\documentclass[
    ,final            
  ]
  {aipproc}

\layoutstyle{6x9}
\usepackage{graphicx}

\begin{document}

\title{Apertif, a focal plane array for the WSRT}

\classification{95.55.Jz, 95.85.Bh, 98.62.Ai, 98.62.Py
}
\keywords      {focal plane array, wide area surveys}

\author{M.~A.~W. Verheijen}{
  address={Kapteyn Astronomical Institute, University of Groningen}
}

\author{T.~A. Oosterloo}{
  address={ASTRON, Dwingeloo}
  ,altaddress={Kapteyn Astronomical Institute, University of Groningen} 
}

\author{W.~A. van Cappellen}{
  address={ASTRON, Dwingeloo}
}
\author{L. Bakker}{
  address={ASTRON, Dwingeloo}
}
\author{M.~V. Ivashina}{
  address={ASTRON, Dwingeloo}
}

\author{J.~M. van der Hulst}{
  address={Kapteyn Astronomical Institute, University of Groningen}
}

\begin{abstract}
  In this paper we describe a focal plane array (FPA) prototype, based
  on Vivaldi elements, developed for the Westerbork Synthesis Radio
  Telescope (WSRT) to increase its instantaneous field of view by a
  factor 25 and double its current bandwidth. This prototype is the
  first step in a project that has the ambition to equip most of the
  WSRT antennas with FPAs to improve the survey speed of the
  telescope. Examples of scientific applications are surveys of the
  northern sky in polarised continuum and HI emission, and efficient
  searches for pulsars and transients.
\end{abstract}

\maketitle


\section{Introduction}

Present day synthesis radio telescopes have limited survey
capabilities because of field of view restrictions. On single dish
radio telescopes this restriction has been alleviated by constructing
multiple feeds which provide several beams on the sky, such as the
succesful Parkes multi-beam system which surveyed the entire southern
sky in HI (Staveley-Smith et al. 1995, 1996),
and the
Arecibo multi-beam system AlfAlfa (Giovanelli et al. 2005).
An alternative method to form multiple beams on the sky is to employ
phased array technology in the focal plane of a radio telescope. The
advantage of using this technique is that there is great flexibility
in forming beams on the sky. Especially in telescopes with small f/D
ratios, it is the only way to form beams on the sky which are closely
packed, touching at their half power points. In addition it is much
easier to optimise the illumination of the telescope dish for maximum
gain or minimum sidelobes.  APERTIF (``APERture Tile In Focus'') is
such a system that is being developed for the Westerbork Synthesis
Radio Telescope (WSRT). In this paper we will briefly describe the
technical details of the focal plane array system under development,
report first results with a prototype and give a brief overview of a
few scientific projects which will come within reach when the WSRT is
equipped with FPA feeds.

\section{The APERTIF system and its prototype}

The development of an FPA system builds upon many years of experience
with phased arrays at ASTRON, an effort motivated by enabling this
technology for the Square Kilometre Array (Kant et al. 2005, Bij de
Vaate et al. 2002).
Design specifications for the APERTIF FPA system are that it should
operate in the frequency range 1000 - 1750 MHz, with a useful
bandwidth of 300 MHz, a system temperature of $\sim 55$ K and an
aperture efficieny of 75\%. For the WSRT this amounts to an effective
area to system temperature ratio of $A_{eff}/T_{sys} \sim 100$
m$^2$/K, if all 14 telescopes are equipped with FPAs.  The goal is to
have 25 beams on the sky or an effective field of view (FoV)
of 8 square degrees. This would increases the survey speed of the WSRT
with a factor of $\sim 20$.

\begin{figure}[tpb]
  \includegraphics[height=.40\textheight]{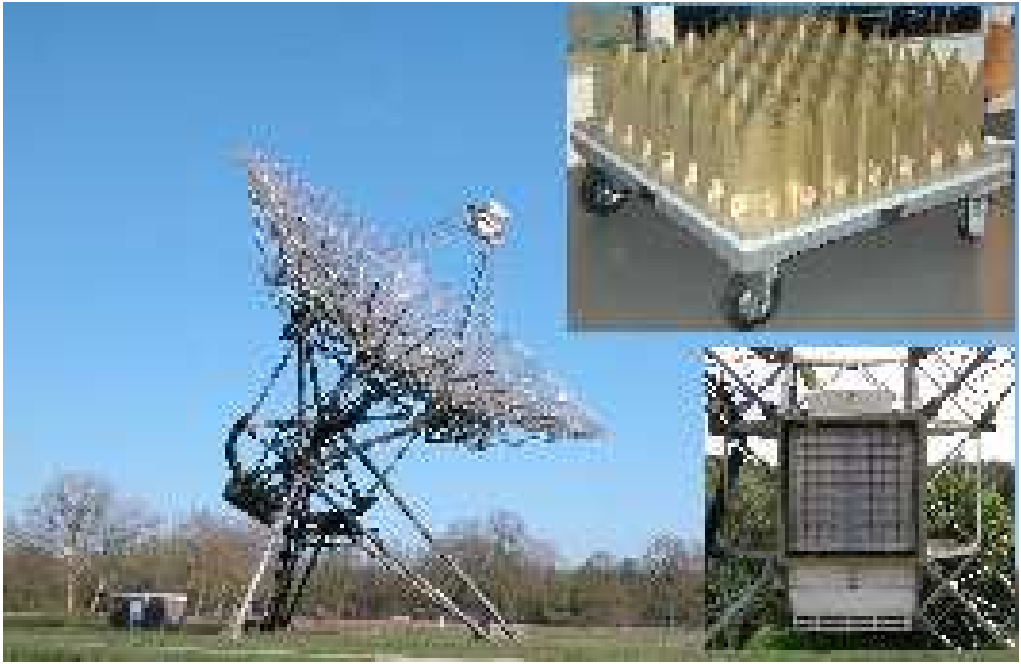}
  \caption{The WSRT telescope RT5 equipped with the APERTIF prototype.
Shown in the inserts is the Vivaldi array in the lab (top right) and
in the focal plane of the telescope (bottom right).}

\end{figure}

A prototype consisting of a dual-polarised Vivaldi array of $8 \times
7 \times 2$ elements has been completed and installed in the focal
plane of one of the WSRT 25-m telescopes. The spacing between the
Vivaldi elements is 10 cm ($\lambda/2$ at 1500 MHz). With this
arrangement one is very close to reaching the ideal situation where
the element beams overlap at their -3dB points while obtaining a dish
illumination with minimal spillover (Ivashina et al. 2007).
Each Vivaldi element has its own LNA. The data is sampled at 12 bits
and the beamforming is all digital. A close up of this prototype in
the focal plane of one of the WSRT telescopes is shown in Figure 1
(bottom right insert).

\begin{figure}[tpb]
  \includegraphics[height=.32\textheight]{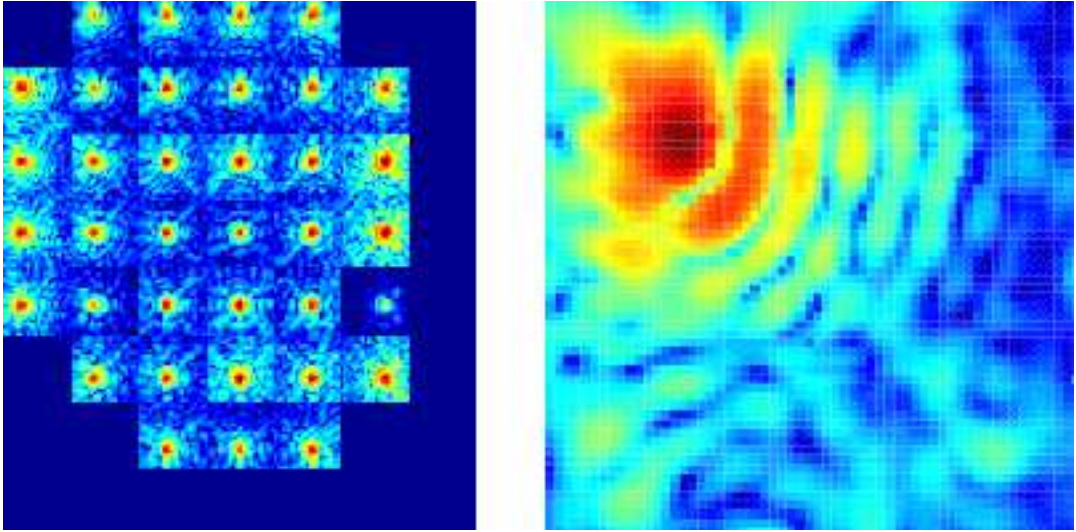}
  \caption{{\bf Left:} Response patterns of the 36 individual Vivaldi
  elements of the FPA that are equiped with an LNA. {\bf Right:} Blow
  up of one of the outer beams to show better the detailed
  response. The effect of coma, due to the fast F-ratio of the dish, is
  obvious.}
\end{figure}

The first observations on sky were made in September 2007. Figure 2
(right panel) shows the response pattern of the individual elements
obtained from observations of the geostationary satellite Afristar at
1490 MHz. Not all elements had been equipped with an LNA, so only 36
of the 56 single-polarisation elements show a response. One can
clearly see the effects of coma when moving away from the centre of
the focal plane (see also the expanded view of the response of an
outer element in the left panel).  Signals from the individual Vivaldi
elements are weighted and combined to digitally form a compound beam,
optimised as much as possible to minimise sidelobes while maximising
the signal to noise ratio and the overall quality of each
beam. Eventually, a proper weighting can also minimise the effects of
instrumental polarisation which will be significant as the X- and
Y-elements are not co-located. A weighting scheme for maximum
signal-to-noise for a source at the field center has been described by
Lo et al. (1966) and Ivashina (2008).
The signals from the Vivaldi elemens can be split after amplification
and used to simultaneously construct multiple compound beams at
different locations on the sky (see Figure 3). %
Figure 3 shows such a set of multiple 
compound beams. Note the flat response over an area of $\sim$ 1.5 degrees. 

\section{First `Light'}

The first astronomical observations with the APERTIF prototype were
performed on a number of sources; Cygnus A, HI in the Milky Way and HI
in M~31. Especially the latter is an excellent demonstration of the
power of the FPA system and is shown in Figure 4.  The right panel
shows a 163 pointing mosaic of M~31 obtained by Braun using the
existing WSRT with a single feed (2008, see also Thilker at al. 2005).
The left panel shows a single pointing with the APERTIF prototype on a
single telescope.  A total of 11x11 compound beams were formed with
the Vivaldi signals of a single telescope pointing. The implication is
that when all WSRT telescopes are equipped with a FPA system, the 163
pointing observation can be carried out in only one or a few
pointings, demonstrating the impressive increase in survey speed.

\begin{figure} [tpb]
  \includegraphics[width=\textwidth]{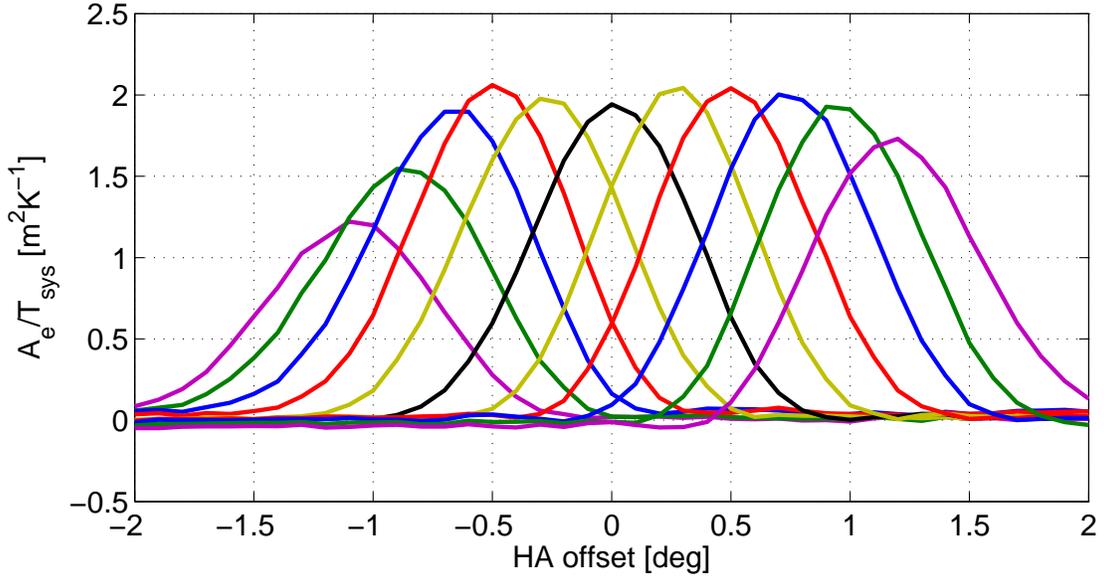}
  \caption{Various compound beams as constructed by combining the
  weighted signals from individual Vivaldi elements. Towards the
  right-side edge of the Vivaldi array (HA offset = +1.5 deg), the
  gain drops off somewhat. On the left hand side, the drop-off is
  more severe because some Vivaldi elements were not equiped with an
  LNA on that side.}
\end{figure}

Figure 5 shows a $3 \times 3$ pointing observation of M~31 with the
APERTIF prototype, covering a $6.5^{\circ} \times 6.5^{\circ}$ field
of view.  The total integration time for this observation was 3
minutes, or 3$\times$6.7 seconds per pointing.  Note in particular the
position velocity diagram along the major axis, illustrating the
behaviour of the rotation curve of M~31, rising fast in the central
region and then dropping to a constant rotation level in the outer
parts. The horizontal band of emission at the lower velocity (higher
frequency) side of the diagram is due to foreground emission from the
Milky Way, filling the entire field of view.

\section{Future prospects}

The goal of the APERTIF project is to equip all 14 telescopes of the
WSRT with a 64 dual polarisation element FPA to produce $5 \times 5$ beams 
on the sky overlapping at half power, thus providing a $3^{\circ} \times
3^{\circ}$ instantaneous field of view.

Once this 25-fold increase in field of view has been realised, the
telescope will have a survey speed (defined as ($(A_{eff}/T_{sys})^2$
FoV BW) of about $15 \times 10^6$ m$^4$K$^{-2}$deg$^2$MHz. One major
programme to undertake is to survey that part of the northern sky that
has been covered by the SDSS in HI and continuum over the 300 MHz
bandwidth, providing the equivalent of the SDSS at radio wavelengths.
Such a survey will require about three years of observing time and
will detect some $2 \times 10^5$ galaxies in HI and even more in radio
continuum. Figure 6 shows the distribution over HI masses in such a
survey (left panel) and the overlap in redshift space with the SDSS
galaxies (right panel). A galaxy with $M_{HI}^*$ will be detectable
out to a redshift of $z = 0.08$ ($v_{hel} = 24,000 km s^{-1}$) in a
single 12 hour observation.

With such a survey many questions regarding the formation and
evolution of galaxies can be addressed: does the local and global
environments of galaxies regulate their gas content? In particular,
how does the HI mass function depend on environment? Current data do
suggest that the low mass slope of the HI mass function steepens
toward higher density regions (Zwaan et al. 2005) 
but the evidence is based on a very small number of objects and
Springob et al. (2005) find contrary results. How does the HI gas
content of galaxies evolve over cosmic time? Although APERTIF will not
be able to observe HI beyond z=0.15, it might be just far enough to
see a hint of cosmic evolution.

What gas depletion mechanisms, besides star formation, play a role and
which mechanism dominates in what environment? Is it correct that in
the denser environments, especially in dense clusters, ram pressure
stripping plays a much more dominant role as compared to the field
where tidal interactions may be more efficient?

How do blue starforming galaxies replenish their gas reservoirs to
sustain their star formation rates? What is the role and fate of the
gas when galaxies evolve from the {\it blue cloud} to the {\it red
  sequence} and what is the residual gas content of galaxies on the
{\it red sequence} (Bell et al. 2004)?

What is the distribution of angular momentum among disk galaxies and
how does this depend on environment? Is there an alignment of galaxy
spin vectors and if so, how does this relate to the large scale
structure in which galaxies are embedded?

All these question are relevant for obtaining a full picture of the
formation and evolution of galaxies with cosmic time, especially since
an HI survey is the only way to learn about the evolution of the gas
content of galaxies.

\begin{figure} [tpb]
  \includegraphics[height=.36\textheight]{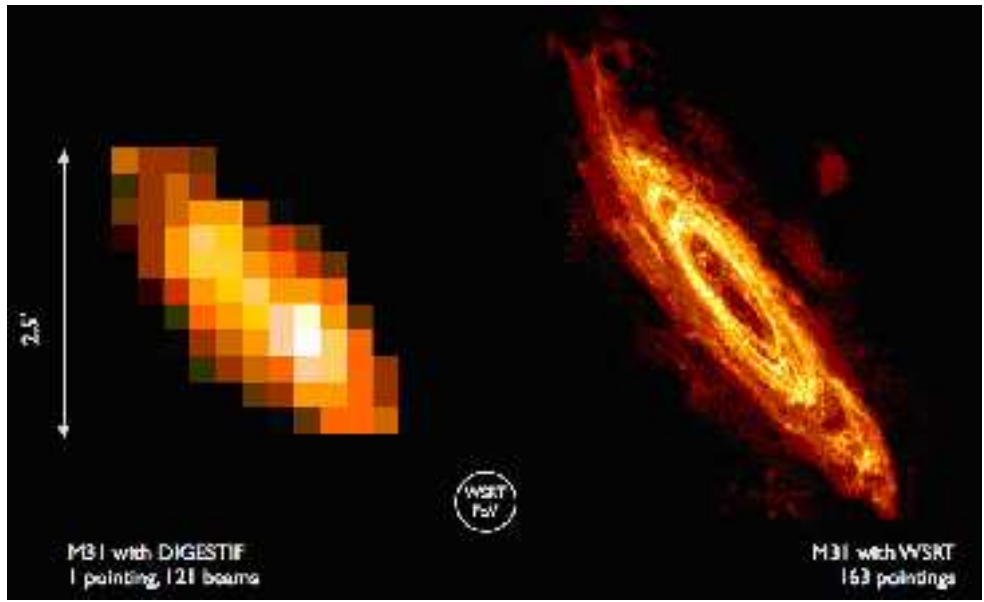}
  \caption{M~31 observed with the WSRT. {\bf Left:} A single-pointing
  measurement with the APERTIF prototype, yielding the resolution of a
  single dish. {\bf Right:} A 163-pointing observation in
  interferometric mode, yielding the full resolution of the 3km
  array. Once all WSRT antennas are equiped with an FPA, the same high
  resolution can be achieved over the full field of view of APERTIF.}
\end{figure}

Such an HI survey would also provide data for several millions of
continuum sources and be fully complementary to the surveys at low
frequency that will be carried out by LOFAR. The radio continuum of
galaxies with either optical or HI redshifts can be used to determine
the star formation rates unambiguously, independent of the effects of
extinction, and hence give an unbiased census of the star formation
density with redshift. From earlier results it is well known that the
star formation density in the universe declined drastically between $z
= 1$ and $z = 0$ (Heavens et al. 2004) but can this be confirmed?
Furthermore, performing such an HI survey in full polarisation mode
would also provide a dense grid of rotation measures from which the
magneto-ionic properties of the Galactic ISM can be probed.

\begin{figure}[tpb]
  \includegraphics[width=\textwidth]{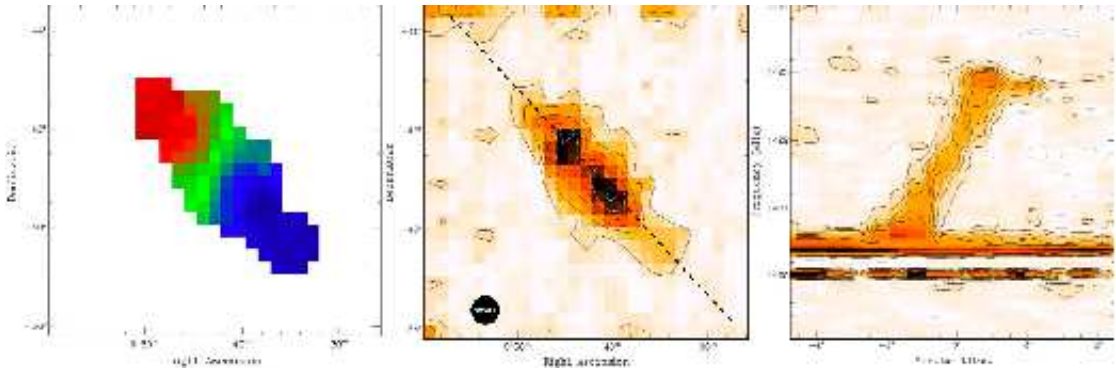}
  \caption{A 3$\times$3 pointing observation with the APERTIF
  prototype of M~31. {\bf Left:} HI velocity field. {\bf Middle:}
  Integrated columns density map. The black circle indicates the FWHM
  of a single dish. {\bf Right:} Position velocity diagram along the
  major axis as indicated in the middle panel. The lower horizontal
  stripe is caused by internally generated RFI which is currently
  removed from the system. The upper horizontal band is Galactic HI
  emission over the full field of view.}
\end{figure}

With the APERTIF system, the WSRT becomes an extremely efficient
pulsar search facility. Thanks to its difraction properties as a
regular east-west array of antennas, the WSRT will become an ideal
instrument to survey the galactic plane for pulsars and obtain precise
positions, periods and period derivatives for several thousands of
pulsars out to 20 kpc. Such a survey not only provides a good census
of the distribution of pulsars, i.e. stellar evolution remnants in the
Galaxy, but also enhances the probability of detecting the occasional
rare object, such as binary pulsars and other exotic objects. With
several thousands of pulsars one will also have access to an equal
number of sightlines through the galaxy enabling studies of the
interstellar medium through its dispersive properties apparent in the
pulsar measurements.

Finally, the large instantaneous field of view of the APERTIF system
on the WSRT will greatly enhance Westerbork's sensitivity to transient
fenomena.

\begin{figure}[tpb]
  \includegraphics[height=.25\textheight]{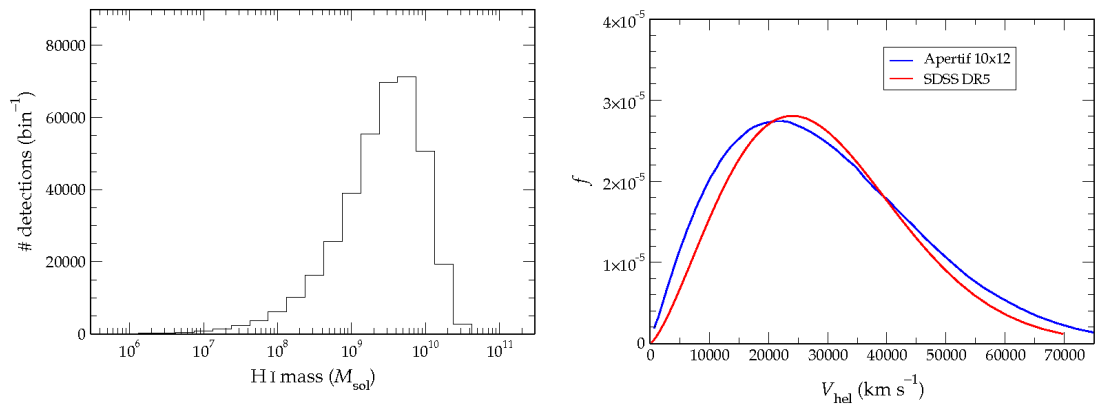}
  \caption{Predicted detection rate (left panel) and redshift distribution
(right panel) of a survey of the SDSS area with the WSRT and APERTIF.}
\end{figure}

\bibliographystyle{aipproc}   

\bibliography{sample}


\end{document}